\documentclass[preprint,aps ,nofootinbib]{revtex4}
\pdfoutput=1
\usepackage[colorlinks=true,linkcolor=blue,urlcolor=blue,filecolor=black,citecolor=red,pdfstartview=FitV,pdftitle={},pdfsubject={},pdfkeywords={},pdfpagemode=None,bookmarksopen=true]{hyperref}
\usepackage[T1]{fontenc} 
\usepackage[figuresright]{rotating}
\usepackage{multirow}
\usepackage{lscape}
\usepackage{subfigure}
\usepackage{graphics}
\usepackage{amsmath}
\usepackage{amsfonts}
\usepackage{amssymb,ulem}
\usepackage[cmyk]{xcolor}
\usepackage{booktabs}

\begin{document}

\title{\boldmath Learning the Black Hole Metric From Holographic Conductivity}
\author{Kai Li $^{1,2,3}$}
\email{lik@ihep.ac.cn}
\author{Yi Ling $^{2,3}$}
\email{lingy@ihep.ac.cn}
\author{Peng Liu $^{4}$}
\email{phylp@email.jnu.edu.cn}
\author{Meng-He Wu$^{5,6}$}
\email{mhwu@sues.edu.cn} \affiliation{$^1$ Kuang Yaming Honors School, Nanjing University, Nanjing 210023, China \\ $^2$
Institute of High Energy
Physics, Chinese Academy of Sciences, Beijing 100049, China\\ $^3$
School of Physics, University of Chinese Academy of Sciences,
Beijing 100049, China \\ $^4$ Department of Physics and Siyuan Laboratory,
Jinan University, Guangzhou 510632, China\\
$^5$ School of Mathematics, Physics and Statistics, Shanghai University of Engineering Science, Shanghai 201620, China \\
$^6$ Center of Application and Research of Computational Physics, Shanghai University of Engineering Science, Shanghai 201620, China}

\begin{abstract}
We construct a neural network to learn the RN-AdS black hole metric based on the data of optical conductivity by holography. The linear perturbative equation for the Maxwell field is rewritten in terms of the optical conductivity such that the neural network is constructed based on the discretization of this differential equation. In contrast to all previous models in AdS/DL (deep learning) duality, the derivative of the metric function appears in the equation of motion and we propose distinct finite difference methods to discretize this function. The notion of the reduced conductivity is also proposed to avoid the divergence of the optical conductivity near the horizon.The dependence of the training outcomes on the location of the cutoff, the temperature as well as the frequency range is investigated in detail. 
This work provides a concrete example for the reconstruction of the bulk geometry with the given data on the boundary by deep learning. 
\end{abstract}
\maketitle

\section{Introduction}\label{sec:intro}

AdS/CFT correspondence as a typical implementation of gauge/gravity duality reveals the deep connections between a $(d+1)$-dimensional gravity theory and a $d$-dimensional quantum field theory \cite{tHooft:1993dmi,Maldacena:1997re,Witten:1998qj,Aharony:1999ti}. In particular, due to the feature of strong/weak duality in large N limit, AdS/CFT correspondence has been proven to be a powerful tool to study the strongly correlated physics via classical gravitational theories. When applying to many-body systems in condensed matter physics, it has formed an important subject which now is dubbed as AdS/CMT duality \cite{Gubser:2008px,Hartnoll:2008vx,Hartnoll:2009sz,McGreevy:2009xe,Hartnoll:2016apf,zaanen2015,ammon2015}. In this field, the traditional way is properly setting the structure of bulk geometry, and then deriving the properties of the many-body system living on the boundary by solving the equations of motion in the bulk with the use of the holographic dictionary. Surprisingly, one finds some of the properties simulate the transport behavior of a strongly coupled system which has been observed in laboratory but is very hard to understand based on the standard perturbative method in quantum field theory \cite{Hartnoll:2008vx,Hartnoll:2009ns,Horowitz:2010gk,Cubrovic:2009ye,Iqbal:2011ae,Horowitz:2013jaa,Ling:2014saa,Ling:2016wuy,Ling:2017jik,Ling:2017naw,Ling:2019gjy,Ling:2020qdd}. In this direction, remarkable progress has been made in understanding the longstanding problems in condensed matter physics by holography, such as the mechanism of high-temperature superconductivity and the non-Fermi liquid behavior of strange metals \cite{Sachdev:2011wg,Zaanen:2018edk}. Nevertheless, this route contains a vital limitation that prevents us from thoroughly solving the problems faced by experimental physicists in laboratory. That is, the derived properties of the dual system greatly depend on the specific structure of spacetime in the bulk. Once the setup of bulk geometry is given, then the transport property of the dual system on the boundary is determined. Usually, based on the consideration of symmetry one may construct the bulk geometry with essential ingredients to observe the expected phenomenon for a dual system, but this is not the general case. In general, we do not know what kind of bulk geometry would give rise to the specific property of the boundary theory as one expects. For instance, there are two fundamental problems which have not been solved by AdS/CMT duality. One is to reproduce the expected phase diagram for the high-temperature superconductivity, the other is to reproduce all the observed features of strange metal in a single holographic model. In this situation, we are facing the inverse problem of the traditional AdS/CMT method: Given the observed features of practical materials in lab, can we construct a holographic model to reproduce such features properly and finally provide a theoretical understanding on these features?  Or more generally, in the context of holography, given the data on the boundary, how can we reconstruct the geometry of the bulk?

Without doubt, this inverse problem is much harder. A lot of efforts have been made to learn the bulk geometry by machine learning/deep learning(DL), which is now viewed as the most advanced technique in artificial intelligence \cite{Hayden:2016cfa,Swingle:2009bg,Swingle:2012wq,Beny(2013),restricted Boltzmann machines,P.Mehta,Carleo:2019ptp,Ruehle:2020jrk,Gan:2017nyt,Hashimoto:2018ftp,Hashimoto:2018bnb,Hashimoto:2019bih,Hashimoto:2021ihd,Tan:2019czc,Akutagawa:2020yeo,Yan:2020wcd,Hashimoto:2020jug,Lam:2021ugb,Hu:2019nea,You:2017guh,Comsa:2019rcz,Krishnan:2020sfg,Bao:2022rup}. As far as we know,  one may classify the current application of deep learning to holography into two categories by the types of boundary data used. One is taking the entanglement on the boundary as the input data \cite{You:2017guh,Lam:2021ugb} and the other is taking the vacuum expectation value (VEV) as the data \cite{Hashimoto:2018ftp,Hashimoto:2018bnb,Tan:2019czc,Yan:2020wcd}.  For the first category, one usually constructs a tensor network as the discretized version of AdS/CFT correspondence, and then transfers this  network to a Boltzmann machine, and a method called entanglement feature learning (EFL) is suggested to learn the spatial geometry from the feature of entanglement on the boundary \cite{You:2017guh}. A generic neural network is also proposed to recover the geometry fluctuation from the multi-region entanglement entropy on the boundary \cite{Lam:2021ugb}. The second category is less ambitious but more relevant to AdS/CMT duality. One just specifies the metric of the bulk to be some simple form with one or more unknown functions, and then constructs a neural network based on the equations of motion for matter fields. The goal of the neural network is to learn unknown functions in the metric by boundary data which are the VEV of dual operators on the boundary. Now such an approach is called AdS/DL method. As the first step, in \cite{Hashimoto:2018ftp}, it is assumed that only one single function in the spacetime metric is unknown. With this ansatz, one attempted to construct the neural network based on the equations of motion of a scalar field and then train the neural network to learn the corresponding space-time metric by inputting the experimental data of magnetization and external magnetic field as initial data. 
Moreover, this AdS/DL method has also been applied to AdS/QCD duality \cite{Hashimoto:2018bnb, Akutagawa:2020yeo, Hashimoto:2021ihd}. Later, the neural network is replaced by another machine learning algorithm called neural ODE, which can serve the same purpose meanwhile yielding more accurate results \cite{Hashimoto:2020jug}. 

In all above work, the neural network is constructed by considering the perturbations of a scalar field. In \cite{Yan:2020wcd}, the perturbation of the metric tensor has been considered and the neural network reflects the RG flow equation of the shear viscosity. In this paper, aiming to apply  AdS/DL method to AdS/CMT duality, we intend to extend the setup to investigate the perturbations of a vector field in the bulk. Specifically, we consider the electromagnetic field $A_{\mu}$ in a charged black hole background and construct the neural network based on the RG flow equation of the optical conductivity of the dual current operator, then train the black hole metric from the data of the optical conductivity on the boundary.

We organize this paper as follows. In section \ref{sec:2}, we derive the equation of motion for the linear perturbation of the Maxwell field and then rewrite this equation with the optical conductivity as the fundamental variable. The neural network is constructed based on the discretized version of this equation. In section \ref{sec:3}, we explain the preparation of the input and output training data. In section \ref{sec:4}, we illustrate the results of deep learning and discuss the effects of the chemical potential $\mu$ and the region of frequency $\omega$ on the final training results. We suggest a novel regularization term to improve training accuracy and save time in hyper-parameter tuning. We also propose several conditions on the regularization term, such as smooth metric and asymptotic AdS, to obtain physically reasonable results. Conclusions and discussions are given in section \ref{sec:5}. Appendix \ref{Appendix: 4 ways of fprime(z) difference} gives details of four kinds of finite difference methods for $f^{\prime}(z)$. Appendix \ref{Appendix: DNN training parameters and results} gives the detailed training methods, hyper-parameters, and all the training results. Appendix \ref{Appendix: The python code} states the running environment of the code.

\section{Building deep neural network by holographic conductivity }\label{sec:2}

In this section, we derive the equation of motion for the optical conductivity in AdS/CMT duality and then construct the corresponding neural network for training the black hole metric. We start with the action of Einstein-Maxwell theory with a negative cosmological constant,
\begin{equation}
    S=\frac{1}{2\kappa^{2}}\int d^{4}x\sqrt{-g}\left(R+\frac{6}{L^{2}} - \frac{F^{ab}F_{ab}}{4}\right),
\end{equation}
where $\kappa^2=8\pi G$ with $G$ the Newton constant, and $\frac{6}{L^{2}}$ is the cosmological constant term with $L$ being the AdS radius. The field strength is $F=dA$, where $A$ is the Maxwell field. From this action, the equations of motion can be derived as,
\begin{equation}\label{eq:ein}
    \begin{aligned}
    R_{ab}-\frac{1}{2}g_{ab}R- \frac{3g_{ab}}{L^{2}} &-\left(F_{ac}F_{b}{^c}-\frac{1}{4}g_{ab}F^{2}\right) =0,\\
    &\nabla^{a}F_{ab}=0.
    \end{aligned}
\end{equation}
We consider the following charged black brane solution to \eqref{eq:ein} with spatially planar symmetry which is also called the AdS-RN metric\footnote{For concreteness, we fix $\kappa=1$ and $L=1$ throughout this paper.},
\begin{equation}\label{eq:bg2}
    d s ^ { 2 } = \frac { 1 } { z ^ { 2 } } \left[ -f(z)  d t ^ { 2 } + \frac { d z ^ { 2 } } { f(z) } +  d x ^ { 2 } +  d y ^ { 2 } \right], \ \  \  A=\mu(1-z)dt,
\end{equation}
where $z\in [0,1]$ is the radial direction of RG flow and $f(z)\equiv 1 -z^3  - \mu ^ { 2 } z ^ { 3 } / 4+ \mu ^ { 2 } z ^ { 4 } / 4$ is the theoretical result by solving Einstein equations. In this paper, we will treat $f(z)$ as the target function that should be learned by the neural network via data training. Parameter $\mu$ is the chemical potential of the dual system on the boundary. In this metric form, the horizon of black brane locates at $z=1$ while the boundary of spacetime locates at $z=0$. Moreover, the Hawking temperature which is given by $T = \frac{12-\mu^2}{16\pi}$, is identified as the temperature of the dual system in equilibrium. Here, we set $\mu\in(0,\sqrt{12}{]}$ to ensure that the Hawking temperature is positive.   Thus, one may change the temperature of the system by adjusting the value of $\mu$.

Now we consider the optical conductivity of the system following the standard procedure of the linear response theory in holographic gravity \cite{Hartnoll:2009sz,Iqbal:2008by}.
To obtain the optical conductivity, we turn on an electric field along $x$ direction by considering the following linear perturbation in the bulk,
\begin{equation}
    \begin{aligned}
        \delta A_{x}=A_{x}(z)e^{-i \omega t}.
    \end{aligned}
\end{equation}
Plugging it into the Maxwell equation, one obtains the linearized equation of motion for $A_{x}$ as
\begin{equation}\label{eq:deqs2}
    z^4 A''_x(z)+2 z^3 A'(z) -z^2 A'_x(z) B(z)  + C(z) A _ { x } = 0,
\end{equation}
where
\begin{equation}\label{eq:bcv}
    \begin{aligned}
        B(z) & =-\frac{z^2f'(z)}{f(z)}+2z,                        \\
        C(z) & =\frac{z^4\omega^2}{f^2(z)}-\frac{\mu^2z^6}{f(z)}.
    \end{aligned}
\end{equation}
The Green function $\langle J_x J_x \rangle$ can be collected as $-\partial_z A_x {/A_x}$ from holographic dictionary, while the applied electric field associated with $A_x$ reads as $E_x = -\partial_t A_x = i\omega A_x$.
From the Kubo formula in holographic gravity \cite{Hartnoll:2008vx,kubo1}, the optical conductivity can be expressed as
\begin{equation}\label{eq:oc2}
    \sigma ( z,\omega ) = \frac { \partial _ { z } A _ { x } ( z ) } { i \omega A _ { x } ( z ) }.
\end{equation}
Next, we intend to rewrite Eq.\eqref{eq:deqs2} as the differential equation in terms of the optical conductivity. For this purpose, we notice that
\begin{equation}\label{eq:dzs}
    \partial_z \sigma = \frac{i A_x'(z)^2}{\omega  A_x(z)^2}-\frac{i A_x''(z)}{\omega  A_x(z)}.
\end{equation}
Furthermore, dividing \eqref{eq:deqs2} by $A_x(z)$ one obtains,
\begin{equation}\label{eq:deqs3}
    z^4 A''_x(z)/A_x+2 z^3 A'_x(z)/A_x -z^2 B(z)A'_x(z)/A_x   + C(z) = 0.
\end{equation}
With the use of Eq.\eqref{eq:dzs}, we obtain the differential equation for the optical conductivity as 
\begin{equation}\label{eq:sigmaeq}
    z^4 \left(i\omega \sigma'(z) - \omega^2\sigma^2  \right) + i \omega \sigma \left(2z^3-z^2B(z)\right) + C(z) = 0.
\end{equation}
Now the original second-order differential equation with variable $A_x(z)$ becomes a first-order differential equation with variable $\sigma(z)$. We are ready to construct the neural network and train the metric function $f(z)$ based on this equation. First, we discretize equation \eqref{eq:sigmaeq} by evenly sampling along the $z$-axis,
\begin{equation}
    \Delta z=\frac{z_h-z_b}{N-1},\quad\quad z(n)=z_b+n\Delta z,
\end{equation}
where $N$ is the number of layers of the network, while $z_b$ and $z_h$ are the locations of the cutoff on the boundary and the horizon, respectively. In addition, $n \in [0,N-1]$, $n \in\mathbf{Z}$, 
$N \in \mathbf{N}^*$ , such as $z(0)=z_b$ and $z(N-1)=z_h$.

We rewrite equation \eqref{eq:sigmaeq} into the real part and the imaginary part separately as
\begin{equation}\label{eq:diff}
    \begin{aligned}
         & \text{Re}\sigma(z+\Delta z)=\text{Re}\sigma(z)+\Delta z\big[ -\frac{f'(z)}{f(z)}\text{Re}\sigma(z)+2\omega \text{Im}\sigma(z)\text{Re}\sigma(z)\big],                                                                 \\
         & \text{Im}\sigma(z+\Delta z)=\text{Im}\sigma(z)+\Delta z\big[-\frac{f'(z)}{f(z)}\text{Im}\sigma(z)+\frac{\omega}{f^2(z)}-\frac{\mu^2z^2}{\omega f(z)}+\omega (\text{Im}\sigma(z))^2-\omega(\text{Re}\sigma(z))^2\big].
    \end{aligned}
\end{equation}

Next, we are concerned with the boundary conditions on both ends of z-axis. At the horizon, we impose the ingoing boundary condition for $A_x(z)$, which takes the form as,
\begin{equation}\label{eqcs}
    A_x(z)=(1-z)^{-\frac{i\omega}{4\pi T}}a_x(z).
\end{equation}
As a result, the optical conductivity becomes,
\begin{equation}\label{eqcssdc}
    \sigma(z,\omega)=\frac{1}{4\pi T(1-z)}+\frac{\partial_za_x(z)}{i\omega a_x(z)}\equiv \frac{1}{4\pi T(1-z)}+\sigma_r(z,\omega),
\end{equation}
where we have denoted the second term as $\sigma_r(z,\omega)$, and we name it the reduced optical conductivity. It is noticed that with the ingoing boundary condition, the first term of the conductivity becomes divergent near the horizon, which easily sabotages the numerics. Therefore, we intend to treat the reduced optical conductivity as the basic variable for the construction of a deep neural network. The discretized version of the equations of motion for the reduced conductivity is given by,
\begin{equation}\label{eqcsdc}
    \begin{aligned}
        \text{Re}\sigma_r(z+\Delta z)= & \text{Re}\sigma_r(z)+\Delta z\left[ -\frac{f'(z)}{f(z)}(\text{Re}\sigma_r(z)+\frac{1}{4\pi T(1-z)})+2\omega \text{Im}\sigma_r(z)\text{Re}\sigma_r(z)\right.         \\& \left.+ \frac{2\omega}{4\pi T(1-z)}\text{Im}\sigma_r(z)-\frac{1}{4\pi T(1-z)^2}\right],\\
        \text{Im}\sigma_r(z+\Delta z)= & \text{Im}\sigma_r(z)+\Delta z\left[-\frac{f'(z)}{f(z)}\text{Im}\sigma_r(z)-\frac{\omega}{(4\pi T)^2(1-z)^2}-\frac{2\omega}{4\pi T(1-z)}\text{Re} \sigma_r(z)\right. \\ &\left.+\frac{\omega}{f^2(z)}-\frac{\mu^2z^2}{\omega f(z)}+\omega (\text{Im}\sigma_r(z))^2-\omega(\text{Re}\sigma_r(z))^2\right].
    \end{aligned}
\end{equation}

In order to reveal the relation between the above discretized equation and a neural network, 
we convert the equation into a matrix form as below,  
\begin{equation}
\left(\begin{array}{c}
\text{Re}\sigma_r(z+\Delta z)  \\
\text{Im}\sigma_r(z+\Delta z)
\end{array}\right) 
=\mathbf{W}_{2\times 2}
\left(\begin{array}{c}
\text{Re}\sigma_r(z)  \\
\text{Im}\sigma_r(z)
\end{array}\right)+
\vec{\mathbf{b}}.
\end{equation}
where
\begin{equation}
    \mathbf{W}_{2\times 2}=\left(\begin{array}{cc}
1-\Delta z \frac{f'(z)}{f(z)} &\Delta z \frac{2\omega}{4\pi T(1-z)} \\
-\Delta z \frac{2\omega}{4\pi T(1-z)} & 1-\Delta z \frac{f'(z)}{f(z)}
\end{array}\right).
\end{equation}
and
\begin{equation}
\quad \vec{\mathbf{b}}=\Delta z \left(\begin{array}{c}
-\frac{f'(z)}{f(z)}\frac{1}{4\pi T(1-z)}-\frac{1}{4\pi T(1-z)^2}+2\omega \text{Im}\sigma_r(z)\text{Re}\sigma_r(z)\\
-\frac{\omega}{(4\pi T)^2(1-z)^2}+\frac{\omega}{f^2(z)}-\frac{\mu^2z^2}{\omega f(z)}+\omega (\text{Im}\sigma_r(z))^2-\omega(\text{Re}\sigma_r(z))^2
\end{array}\right).
\end{equation}

Here, $\mathbf{W}_{2\times 2}$ that contains the information of the spacetime metric can be regarded as the weight matrix of a neural network. The weight matrix in deep learning represents the connecting parameters of the neurons of adjacent layers. And naturally, $\vec{\mathbf{b}}$ can be regarded as the bias term of a network. In addition, according to the matrix form, the activation function is the identical mapping. 

\begin{figure}[htp]
    \centering
    \includegraphics[width=4.5cm,angle=90]{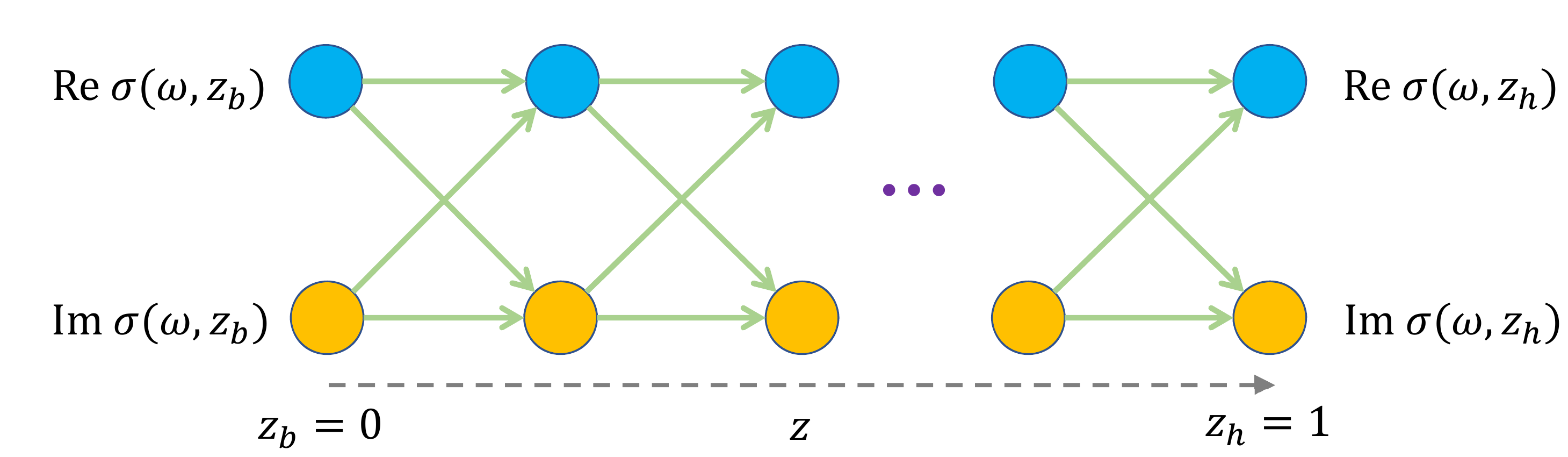}
    \caption{
        The structure of the deep neural network. The holographic optical conductivity propagates from the boundary to the horizon.
    }
    \label{fig:DNN equation}
\end{figure}

As a result, we construct the following neural network to represent the discretized equation of motion for the reduced conductivity (Fig. $\ref{fig:DNN equation}$). 

Physically speaking, we can consider the network structure as the spacetime structure because the connecting weights of the network contain the metric information and the propagation direction of the network is the holographic direction \cite{Hashimoto:2018ftp}. Also, we can imagine a scene where the conductivity travels along the network by perceiving the spacetime information locally.

However, we notice that the derivative of the metric function, namely $f^{\prime}(z)$, appears in the equations of motion as well, which is in contrast to the discretized version of the equations of motion appearing in previous literature on deep learning in holography \cite{Hashimoto:2018ftp,Yan:2020wcd}.
In principle, we may treat $f(z)$ and $f'(z)$ as two independent variables and train them independently. In practice, however, we find that this makes the optimization process of deep learning much more difficult.
To obtain a feasible deep learning process, we discretize $f'(z)$ in terms of its relation to $f(z)$.
There are many different ways to discretize $f'(z)$. Here we select four distinct varieties as listed below and investigate their effects on the final results (see detailed information in Appendix \ref{Appendix: 4 ways of fprime(z) difference}.).
\begin{enumerate}
    \item $\ln f(z)$ forward: $\frac{f^{\prime}(z)}{f(z)}=\big(\ln f(z)\big)^{\prime}\approx \frac{\ln f(z+\Delta z)-\ln f(z)}{\Delta z}$.
    \item $\ln f(z)$ middle: $\frac{f^{\prime}(z)}{f(z)}\approx \frac{\ln f(z+\Delta z)-\ln f(z-\Delta z)}{2\Delta z}$.
    \item $ f(z)$ forward: $f^{\prime}(z)\approx \frac{f(z+\Delta z)- f(z)}{\Delta z}$.
    \item $ f(z)$ middle: $f^{\prime}(z)\approx \frac{f(z+\Delta z)- f(z-\Delta z)}{2\Delta z}$.
\end{enumerate}

We introduce the loss function to evaluate the difference between the true values and the results predicted by the neural network. One criterion of designing a neural network is to make the loss function as small as possible. According to the previous work \cite{Hashimoto:2018ftp,Yan:2020wcd}, we introduce two loss functions as below,
\begin{equation}\label{eqsds}
    L_{\mathrm{total},1}=\frac{1}{N}\sum_{\text {data }}\left|\sigma\left(\omega, z_{h}\right)-\hat{\sigma}\left(\omega, z_{h}\right)\right|+L_{\mathrm{REG}},
\end{equation}
\begin{equation}\label{eqs}
    L_{\mathrm{total},2}=\frac{1}{N}\sum_{\text {data }}\left|\sigma\left(\omega, z_{h}\right)-\hat{\sigma}\left(\omega, z_{h}\right)\right|^2,
\end{equation}
where the regularization term is,
\begin{equation}\label{eqsfc}
    L_{\mathrm{REG}}=c_{1}[f(z(0))-1]^{2}+ \frac{c_{2}}{(n_{\text{epoch}}/10)^{1.5}} \sum_{n=0}^{N-1} \frac{1}{z(n)^2}[f(z(n+1))-f(z(n))]^{2}.
\end{equation}
In the above equations, $\sigma$ is the input data while  $\hat{\sigma}$ is what the network predicts. $c_1$ and $c_2$ are  hyper-parameters, which we can tune manually.
 $n_{\text{epoch}}$ is the number of epochs we run, where an epoch is defined as the period that the full data set propagates through the neural network once.
Here the regularization term $L_{\text{REG}}$ contains two terms and they are used to find a reasonable metric, which differs greatly from the common effect of overcoming the overfitting. The first term is to guarantee the asymptotically AdS property of spacetime at $z_b=0$, while the second term is to suppress the possibility of large gradients to promote the efficiency of the neural network to figure out a smooth metric function numerically. We find both terms are important for the deep learning process, just as in previous work \cite{Hashimoto:2018ftp,Tan:2019czc,Yan:2020wcd}.

\section{The setup for training data and discretization}\label{sec:3}

In this section, we present the setup for the training data and figure out the best way to discretize $f'(z)$. 
Given the theoretical result of the metric function $f(z)$, then from \eqref{eq:sigmaeq} one can numerically obtain the data of optical conductivity from the boundary $(\sigma(z_b))$ to the horizon $(\sigma(z_h))$ for any specified frequency $\omega>0$, as performed in ordinary holographic approach. Now we reverse the problem by setting $f(z)$ as an unknown function and try to learn it by inputting the data of optical conductivity. For this purpose, near the boundary, $z=0$ we fix the location of the cutoff at $z_b = 0.01$  and input 2000 numerical data of optical conductivity with $\omega$ uniformly sampled along $(0.1,1]$ as initial data at the cutoff. Next, since the conductivity becomes divergent at the horizon $z=1$, we also need to introduce a cutoff  $z_h$ near the horizon. Now for each input data, one can generate the data of conductivity at $z_h$ as output data through the neural network. Finally, one can train the neural network to learn the metric function by comparing the output data with the theoretical results. All the training data $\{(\sigma(z_b),\sigma(z_h)) \}$ we use can be found in \cite{dataset}.  Here, we study two cases - $z_h = 0.9$ and $z_h=0.99$ to test the learning ability of the neural network.

Next, we need to fix the number of layers in the neural network. The discretization in the process of deep learning introduces truncation error, which can be decreased by increasing the number of layers in the neural network. In theory, constructing deeper neural networks with more layers would improve the accuracy, however, with the price of consuming time and intensifying resources. In practice, we find that an 11-layer neural network in this work suffices to provide results that are strikingly close to those of deeper networks.

\begin{figure}[htp]
    \centering
    \includegraphics[width=13cm]{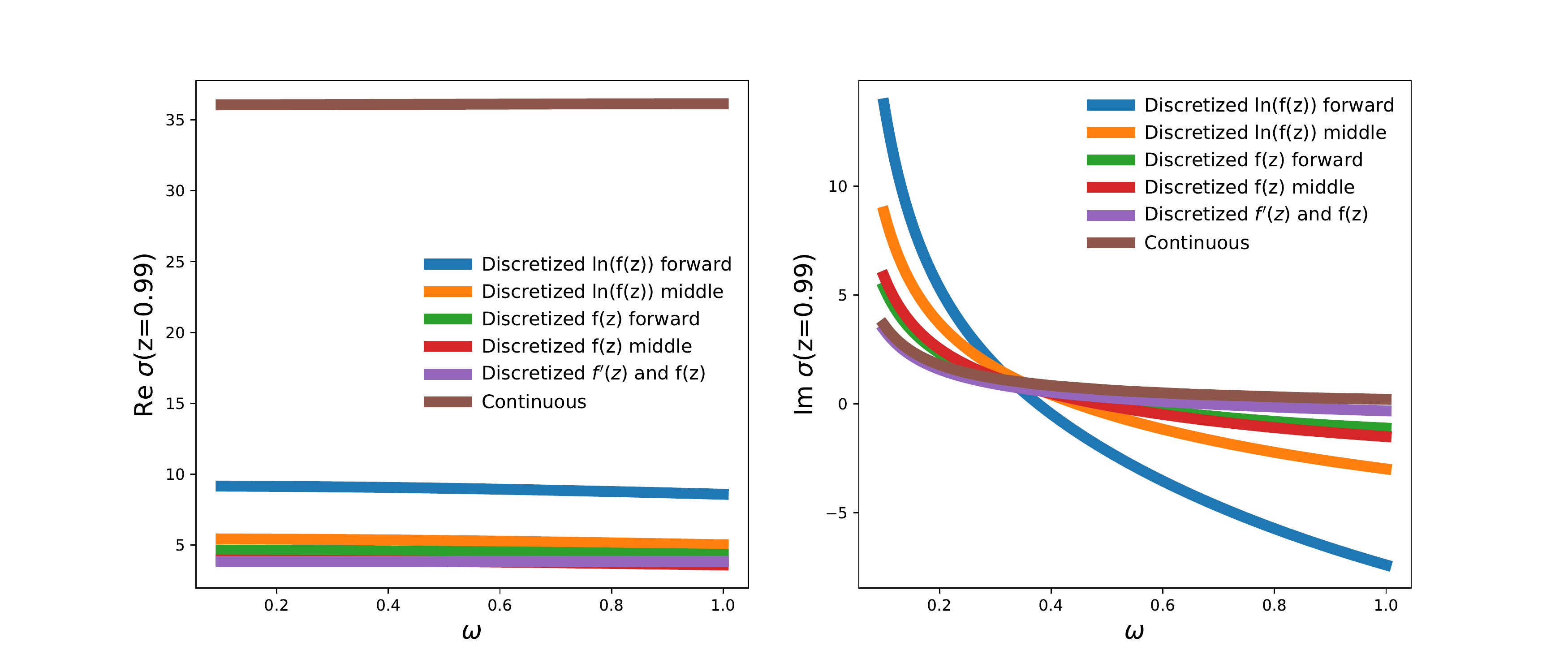}
    \caption{
        The standard conductivity at $z=0.99$ generated by the neural network with different discretizations of  $f^{\prime}(z)$. The left (right) plot is the real (imaginary) part of the optical conductivity.
        The brown curve is the numerical result by directly solving the differential equation with the finite difference method.
    }
    \label{fig:real 0.99}
\end{figure}

\begin{figure}[htp]
    \centering
    \includegraphics[width=13cm]{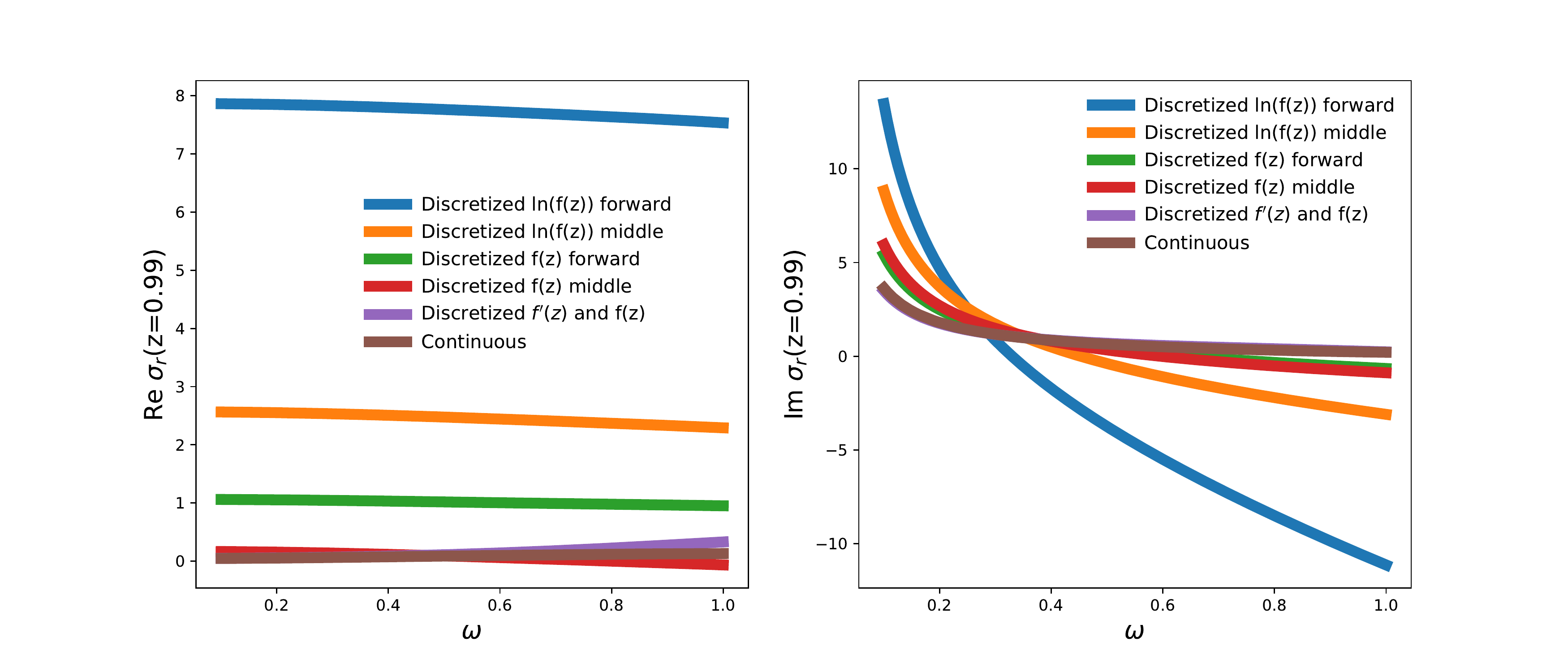}
    \caption{
        The reduced conductivity at $z=0.99$ generated by the neural network with different discretizations of  $f^{\prime}(z)$. The left (right) plot is the real (imaginary) part of the optical conductivity.}
    \label{fig:reduced 0.99}
\end{figure}

Finally, we intend to pick out the best way of discretizing $f'(z)$ for the neural network. 
For this purpose, we show the output data of the standard conductivity and the reduced conductivity at $z_h=0.99$ with $\mu=1$ in Fig. \ref{fig:real 0.99} and Fig. \ref{fig:reduced 0.99}, respectively, which are generated by the neural network with various discretizations of  $f'(z)$. We also present the numerical result by directly solving the differential equation with the finite difference method, which might be viewed as the ``true'' values of the conductivity, namely the data obtained by the deep neural network in the continuous limit.

First, let us focus on the output of the standard conductivity in Fig. \ref{fig:real 0.99}. It is noticed that the data of 
the conductivity obtained by the ``$f(z)$ forward'' discretization looks closer to the data by the continuous limit. However, in practice, we find that the ``$\ln f(z)$ forward'' discretization performs more robustly, and with it one can get more accurate metric information than ``$f(z)$ forward'' discretization (see the comparison in Table \ref{data}). We can make the difference of results between ``$f(z)$ forward'' and ``$\ln f(z)$ forward'' smaller by increasing the number of network layers, but the ``$\ln f(z)$ forward'' method is intrinsically more robust. Therefore, with comprehensive consideration, we decide to adopt ``$\ln f(z)$ forward'' method to train the network. Similarly, we compare the output of the reduced conductivity in Fig. \ref{fig:reduced 0.99}, and find that ``$f(z)$ forward'' discretization is the best way for deep learning process of the reduced conductivity.

We obtain the similar results for the case with  ($z_h=0.99$, $\mu=2$), ($z_h=0.9$, $\mu=2$) and other combinations (in Appendix \ref{Appendix: DNN training parameters and results}). As a result, we choose ``$\ln f(z)$ forward'' discretization for the standard conductivity and ``$f(z)$ forward'' discretization for the reduced conductivity in the construction of the neural network.

More importantly, we find that the output data of the reduced conductivity at $z_h$ is much closer to the data of the continuous limit than that of the standard conductivity. In particular, as $z_h$ approaches the location of the horizon, the reduced conductivity exhibits its advantages more evidently since the divergent part has been peeled off. So in the next section, we focus on the results of the neural network constructed with the reduced conductivity. For full results, please see \cite{dataset}.

\section{Results of the learned metric}\label{sec:4}

\subsection{The result of learned metric with $\mu$=1 and $z_h$=0.99}
\begin{figure}[ht]
     \centering
    \begin{subfigure}
        \centering       \raisebox{0.2\height}{ \includegraphics[width=8cm]{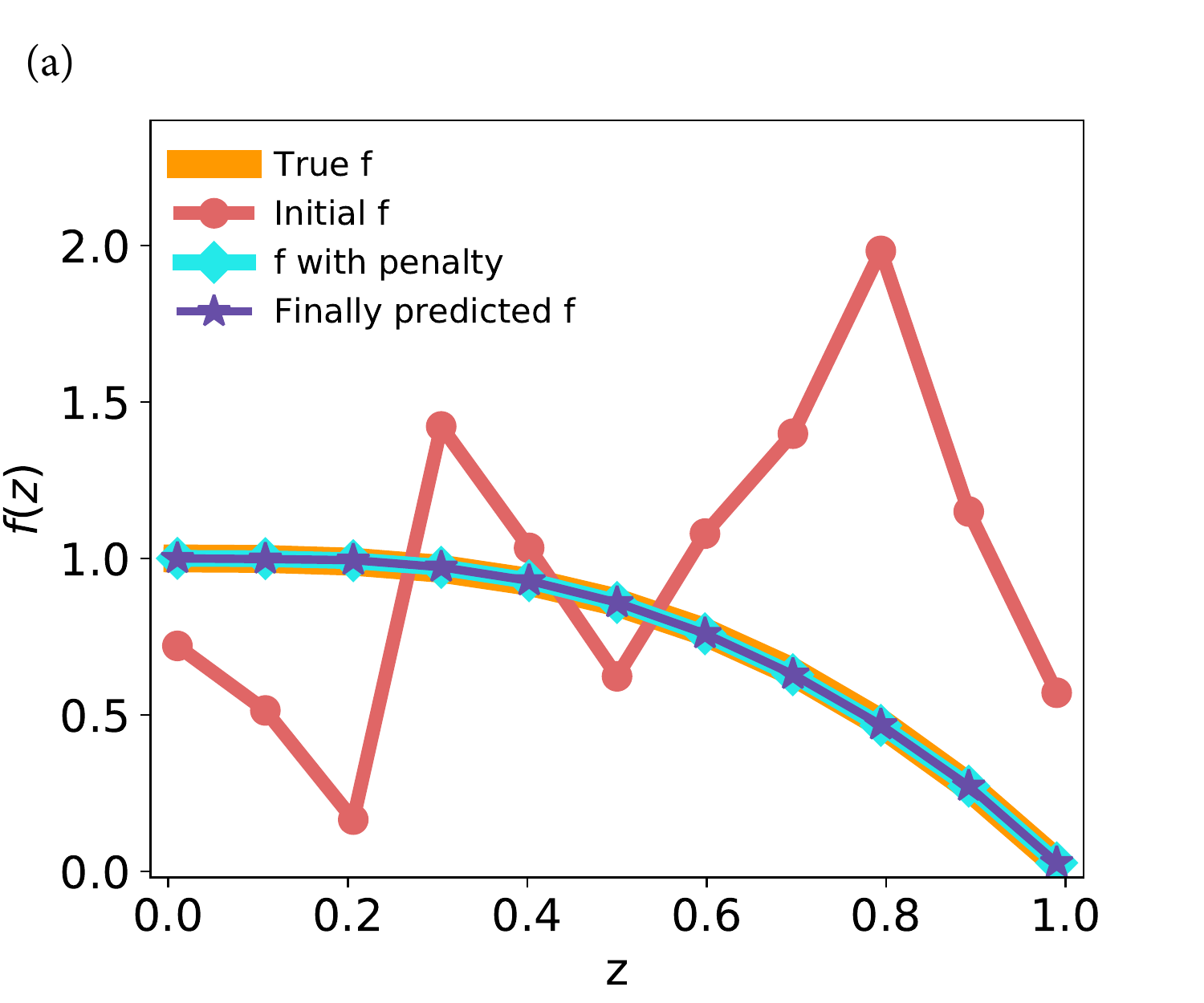}}
    \end{subfigure}
    \quad
      \begin{subfigure}
        \centering
        \includegraphics[width=6cm]{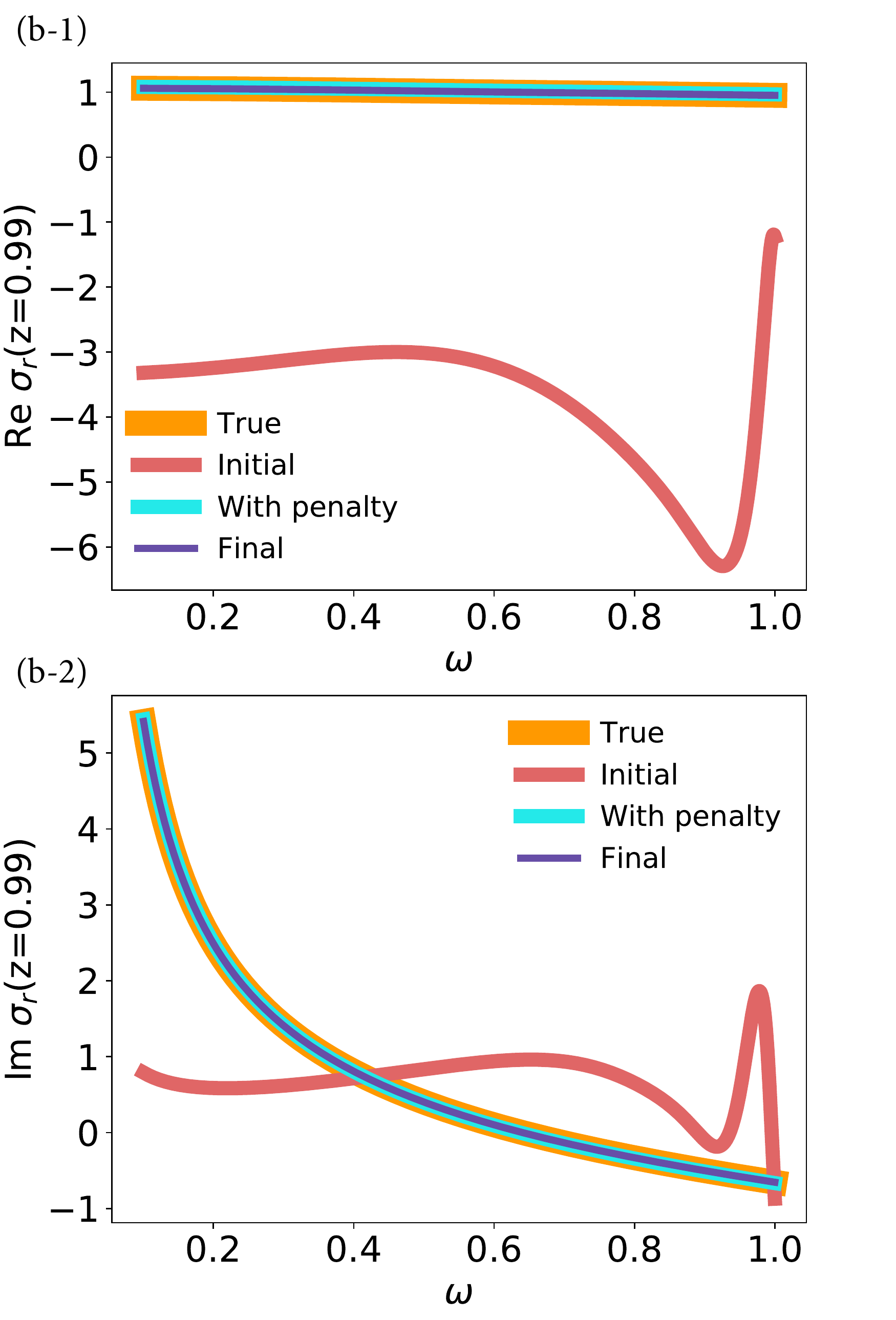}
    \end{subfigure}
    \centering
    \caption{The results of deep learning. (a) The learned metric results. The orange curve represents the true metric $f(z)$. The red curve represents the random initial weights of the metric in the network. The blue curve represents the trained metric after the first training process, and the penalty means the regularization term in the first loss function. The purple one is the final metric after double training procedures. (b-1) The real part of the reduced conductivity generated by different metrics. The orange, red, blue, and purple curves represent the real part of reduced conductivity generated by the true metric, the random initial metric, the metric after the first training process, and the final metric after double training procedures, respectively. (b-2) The imaginary part of the reduced conductivity generated by different metrics.
    }
    \label{fig:result_metric_conductivity}
\end{figure}

Firstly, we show a typical example of the training results for the metric function $f(z)$ with $\mu$=1 and $z_h$=0.99, which is illustrated in Fig. \ref{fig:result_metric_conductivity}. The left figure is the result of the learned metric. It shows that after the deep learning process, the initial randomly selected metric becomes the true metric. Two plots on the right-hand side are the output data of the reduced conductivity at $z_h$. It shows that the reduced conductivity generated by the initial metric is far away from the true one, while after the deep learning process, it is quite close to the true conductivity, indicating that the neural network has successfully learned the metric from the reduced conductivity. Also, we find that after the first training process, the results of both the metric and reduced conductivity are almost as good as the final ones. For more details on the training methods and the training results, please see Appendix \ref{Appendix: DNN training parameters and results}. Because the target of the deep learning is the metric, next we show the results for the metric only.
\subsection{The comparison of results between $\mu$=1 and $\mu$=2 at $z_h$=0.99}
\begin{figure}[htbp]
    \centering
    \subfigure[$\mu$=1]{
        \centering
        \includegraphics[height=2in,width=2.5in]{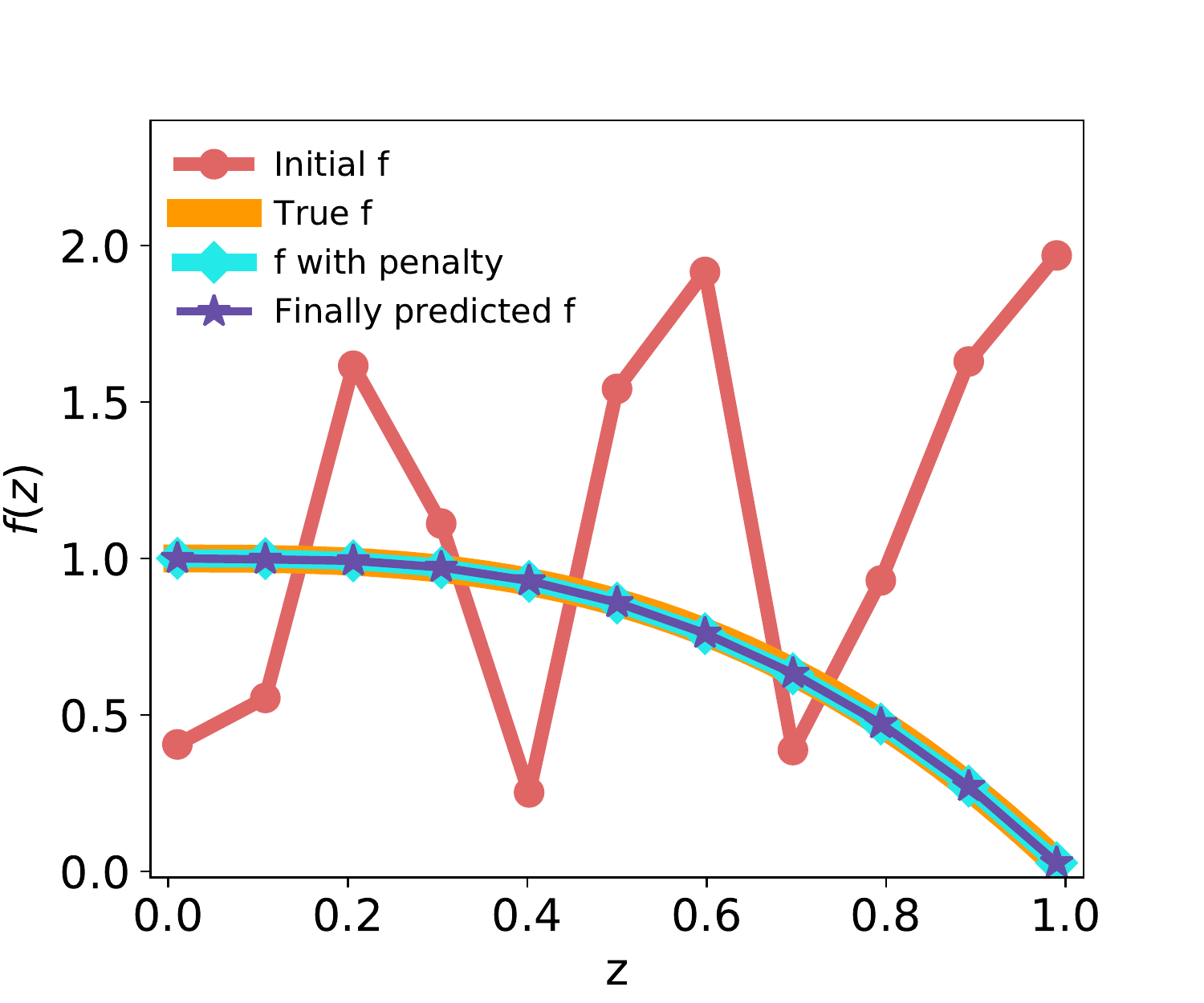}
    }
    \centering
    \subfigure[$\mu$=2]{
        \centering
        \includegraphics[height=2in,width=2.5in]{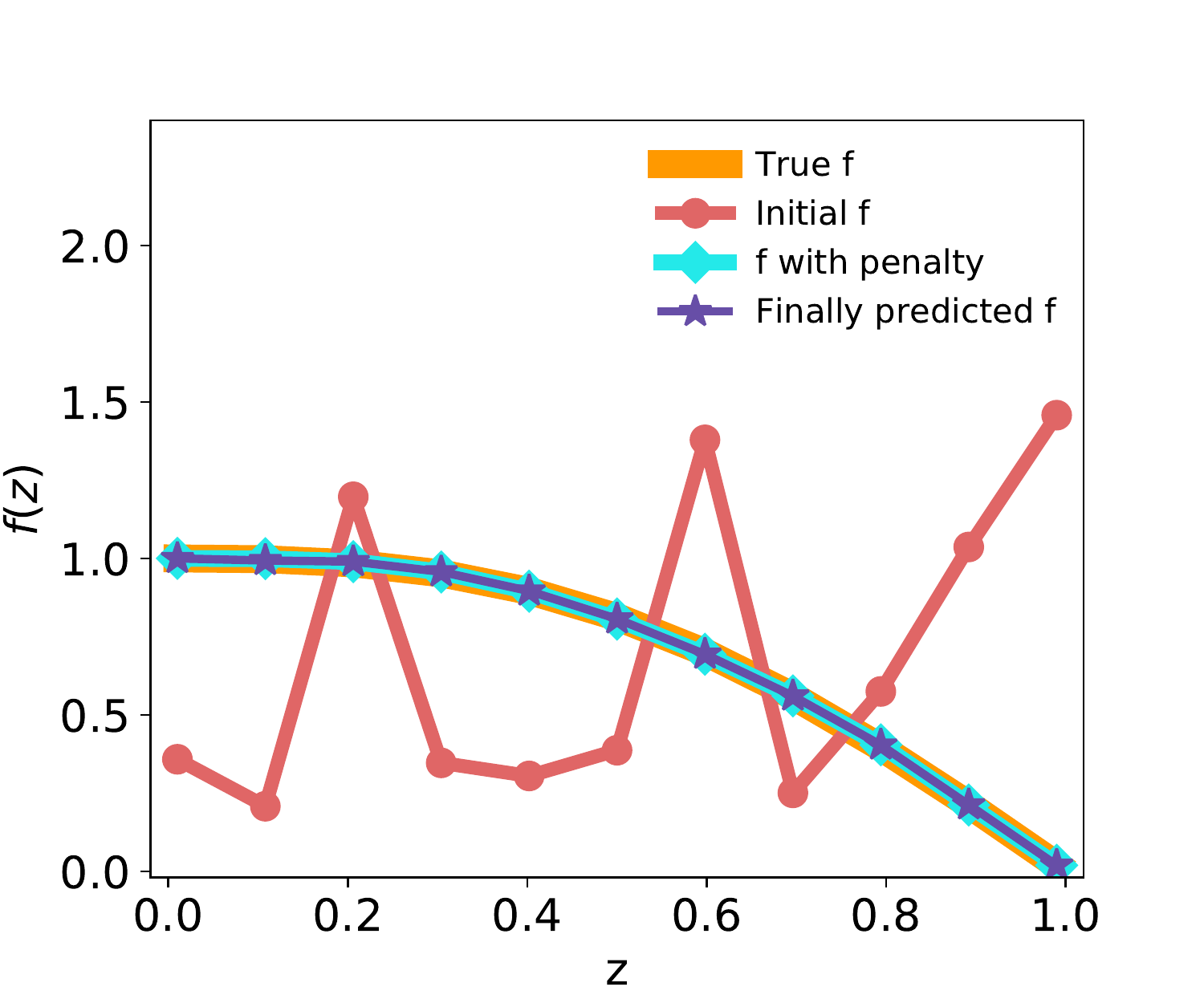}
    }
    \centering
    \caption{
        The DNN training results of the reduced conductivity with different values of chemical potential  ($z_h=0.99,\omega\in(0.1,1]$). 
    }
    \label{fig:reduced_z=0.99_mu=1/2}
\end{figure}

In this subsection, we discuss the effects of the chemical potential $\mu$ on the training results, which can also be understood as the influence of temperature.
Fig. \ref{fig:reduced_z=0.99_mu=1/2} is the deep learning results of the reduced conductivity at $z_h$=0.99 with $\mu=1$ (left plot) and $\mu=2$ (right plot), respectively. We have tried various initial guesses and found they all converge to the true values of $f(z)$ with great accuracy. This shows that the neural network is powerful and robust in learning the metric from optical conductivity.

\begin{figure}[htp]
    \centering
    \includegraphics[width=8cm]{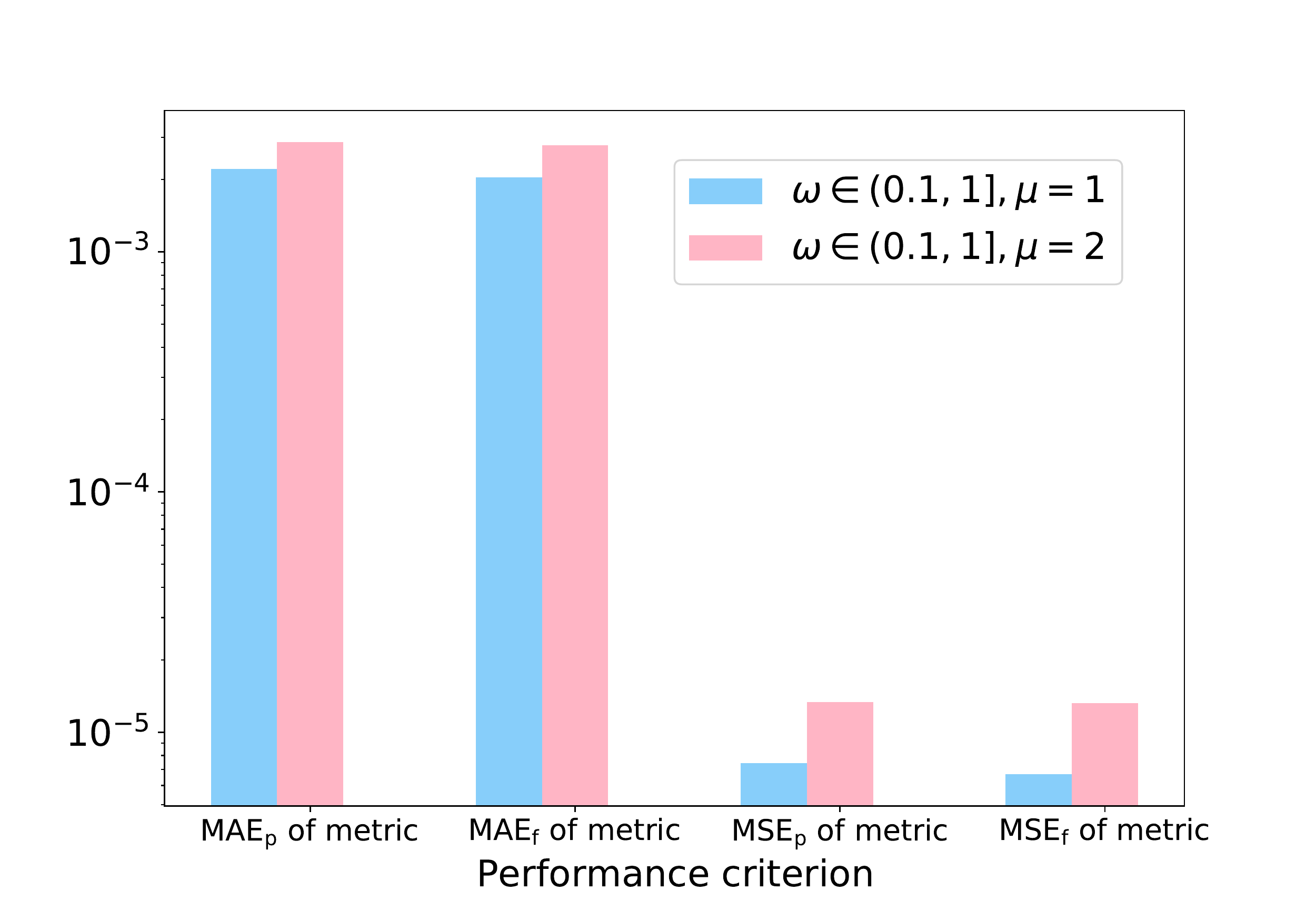}
    \caption{
        The comparison of deep learning results for the reduced conductivity ($z_h=0.99$, $\mu=1,2$). MAE and MSE are two types of criteria functions used to evaluate the performance of training the network, and their definitions may be found in Appendix \ref{Appendix: DNN training parameters and results}. The subscript $p$ represents the penalty and it means the results for the first training procedure. Subscript $f$ represents final and it means the results for the final training procedure.
    }
    \label{fig:comparison z=0.99 mu=1andmu=2 reduced}
\end{figure}

We show more concrete performance criteria in Fig. \ref{fig:comparison z=0.99 mu=1andmu=2 reduced}. It is noticed that the effect of the deep learning for $\mu=1$ is better than that of the case $\mu=2$. This result holds also for many other training data \cite{dataset}. Nevertheless, one can see that both training results are greatly improved after the second training process.

\subsection{The comparison of training results for $\mu$=1 and $z_h$=0.99 with different $\omega$ ranges}
In this subsection, we discuss how the different ranges of the frequency $\omega$ affect the final training results.
Previously, the research on the shear viscosity provided a positive answer to whether the neural network can learn the metric with the data in a narrow range of $\omega$ \cite{Yan:2020wcd}. Here we intend to justify if this is also true for optical conductivity. For this purpose, we study two different ranges of $\omega$: $(0.1,0.11]$ and $(0.99,1]$, each of which takes 2000 data points into account.

\begin{figure}[htbp]
    \centering
    \subfigure[$0.99<\omega\leq 1$]{
        \centering
        \includegraphics[height=2in,width=2.5in]{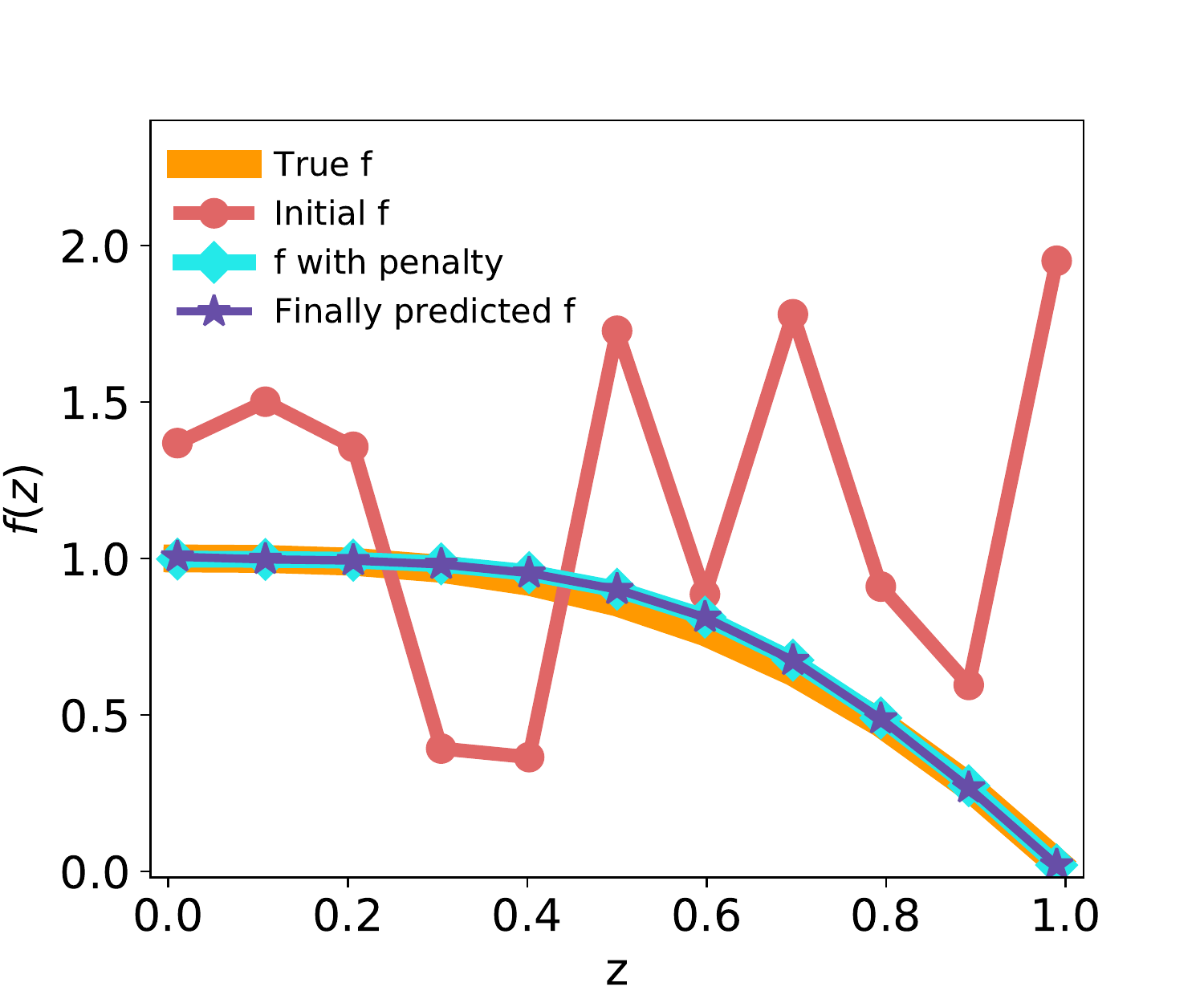}
    }
    \centering
    \subfigure[$0.1<\omega\leq 0.11$]{
        \centering
        \includegraphics[height=2in,width=2.5in]{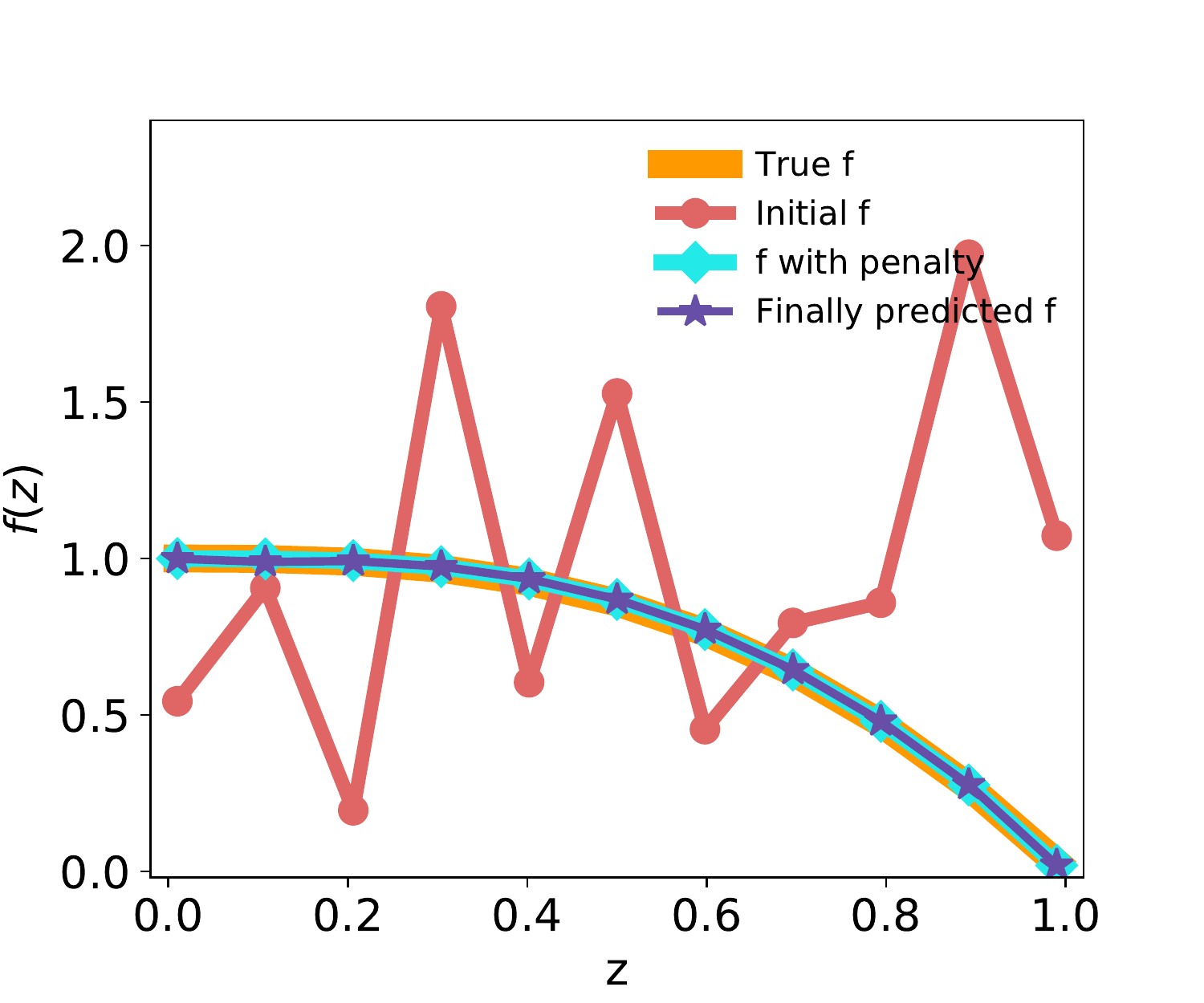}
    }
    \centering
    \caption{The DNN training results of the reduced conductivity at $z_h=0.99$ with $\mu=1$ and $\omega\in (0.99,1]$ (left plot) and $(0.1,0.11]$ (right plot).}
    \label{fig:z=0.99mu1omega=0.99-10.1-0.11comparison}
\end{figure}

\begin{figure}[htp]
    \centering
    \includegraphics[width=8cm]{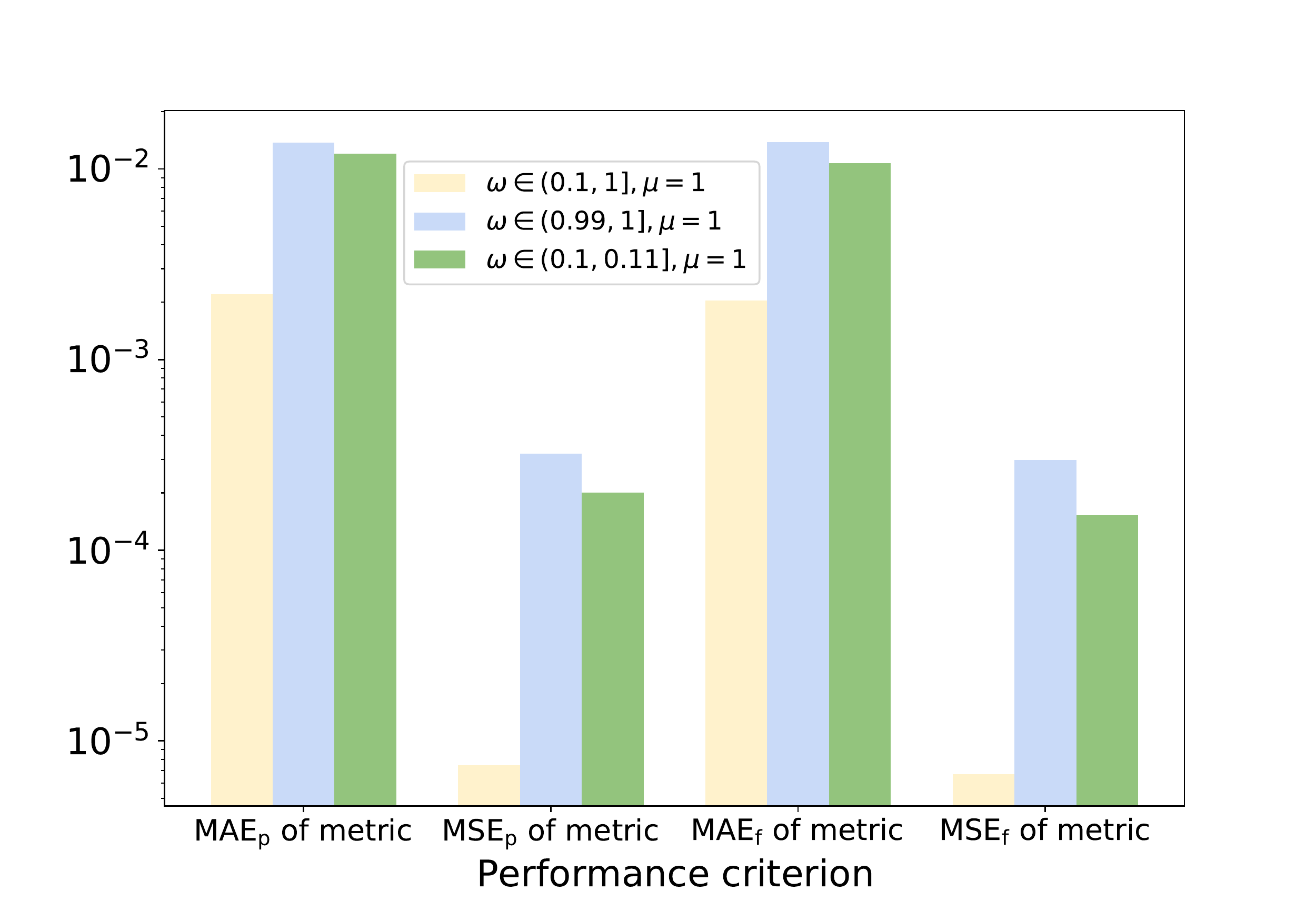}
    \caption{The comparison of deep learning results for the reduced conductivity among three different ranges of $\omega$.}
    \label{fig:criterion_reduced_mu=1_z=0.99.pdf}
\end{figure}

Fig. \ref{fig:z=0.99mu1omega=0.99-10.1-0.11comparison} shows the deep learning results of the reduced conductivity at $z_h=0.99$ with $\mu=1$ and $\omega\in (0.99,1]$ (left plot) and $(0.1,0.11]$ (right plot), respectively. Fig. \ref{fig:criterion_reduced_mu=1_z=0.99.pdf} gives the concrete performance criteria of three different $\omega$ ranges at $z_h=0.99$ of $\mu=1$. We find that the training performance is better when the range of $\omega$ is wider. In addition, a larger $\mu$ will worsen the training outcomes, as illustrated in Appendix \ref{Appendix: DNN training parameters and results}.
\newpage
\section{Conclusion and discussion}\label{sec:5}
We have constructed a neural network to learn the RN-AdS black hole metric in the bulk based on the data of optical conductivity on the boundary by holography. The equation of motion that we recast into a neural network is generated by perturbing the vector field, thus enriching the prior research that only studied the scalar field or metric tensor field. In contrast to previous models, in this circumstance, the derivative of the metric function  $f'(z)$ appears in the equation of motion, and we have proposed four distinct finite difference methods to discretize $f'(z)$. We have investigated their performance during the deep learning process in detail. Furthermore, to recast the equations of motion into a numerically feasible neural network, we have defined the reduced conductivity to avoid the divergence of the optical conductivity near the horizon.
In addition, we have proposed a novel regularization term that automatically tunes the hyper-parameters, which ensures the robustness and efficiency of the training methods.
We have also discussed the dependence of the training outcomes on the location of the cutoff $z_h$, the temperature as well as the frequency range.
It turns out that the network is harder to train as $z_h$ approaches the horizon and as the temperature decreases. 
Given the number of data points, the training results with a wider range of $\omega$ are better than those with a narrower range. This can be understood from the fact that data from wider ranges of frequency contains more information than that from the narrower ranges.

This work has explicitly demonstrated the remarkable power of deep learning in the reconstruction of the spacetime with the given data on the boundary. 
For further study, we expect the AdS/DL method may be applied to AdS/CMT duality and shed light on the open problems in strongly coupled many-body system.  For instance, given the RG flow data of the optical conductivity of the strange metal, the neural network would learn the metric of the bulk geometry which is capable of reproducing all the transport features of the strange metal. Currently, such kind of metric in the framework of AdS/CMT is unknown. Without doubt, the neural network presented in this paper is too simple to accomplish this task. We expect it could be developed into a network with more abundant structure and functions, such that its ability of learning the background information could be greatly improved. As the next step, one could consider a neural network with more neurons such that it could learn more unknown functions rather than a single unknown function in the metric. In addition, we expect the AdS/DL method may be applied to more holographic models and learn the bulk geometry by inputting the data of other transport quantities such as the thermal conductivity, etc. 

 An even more ambitious goal of AdS/DL is to learn the action of the dual gravity system from boundary data, which is of crucial significance not only for finding holographic models to understand important phenomena in dual systems but also for comprehending the implications of machine learning in holographic reconstruction of spacetime geometry. However, many challenges persist in realizing this goal, such as the degeneracy between the boundary data and the action of the dual theory, the construction of a machine learning model that can establish a relationship between the boundary data and the action represented by a symbolic system, etc. Recent advancements in machine learning offer promising prospects for directly learning the action from boundary data. For instance, the representation of symbolic space is comparable to natural language, and there exist highly effective methods, such as seq2seq \cite{sts}, that can effectively address the problem. The SymbolicMathematics \cite{symbolic}, empowered by seq2seq, is even more powerful in solving integral problems than well-known commercial software such as Mathematica and Matlab. These methods provide valuable strategies for representing and exploring the symbolic space. To solve the problem of degeneracy, on one hand, we can reduce the necessary variables in the model based on physical considerations, such as symmetry requirements. On the other hand, compared to the electrical conductivity that we currently consider, one may further reduce the degeneracy by increasing the type of boundary data, such as thermal conductivity, entropy, etc. Furthermore, a crucial capability of machine learning is generalization, meaning that it has the potential to address problems outside its training data range. Currently, advances such as Diffusion models and ChatGPT have robustly demonstrated this \cite{tasks,answers}. We have reason to believe that given sufficiently high-quality data sets, machine learning has the potential to learn more intrinsic properties of holographic gravity and greatly contribute to the development of the AdS/DL.

\section*{Acknowledgments}

We are very grateful to Chen Bai, Yi Gu, Jiahao He, Yu Tian, Xiaoning Wu and Hongbao Zhang for helpful discussions.
This work is supported in part by the Natural Science Foundation
of China under Grant No.~11875053 and 12035016. It is also supported by Beijing Natural Science Foundation under Grant No. 1222031 and by the creative practice training project of UCAS.

\appendix
\section{Four types of finite differences of $f^{\prime}(z)$}
\label{Appendix: 4 ways of fprime(z) difference}
\begin{enumerate}
    \item $\ln f(z)$ forward
          \begin{equation}\label{eq}
              \begin{aligned}
                  \text{Re}\sigma(z+\Delta z)= & \text{Re}\sigma(z)+(\ln f(z)-\ln f(z+\Delta z))\text{Re}\sigma(z)+\Delta z\big[ 2\omega \text{Im}\sigma(z)\text{Re}\sigma(z)\big], \\
                  \text{Im}\sigma(z+\Delta z)= & \text{Im}\sigma(z)+(\ln f(z)-\ln f(z+\Delta z))\text{Im}\sigma(z)+\Delta z[\frac{\omega}{f^2(z)}-\frac{\mu^2z^2}{\omega f(z)}     \\
                                               & +\omega (\text{Im}\sigma(z))^2-\omega(\text{Re}\sigma(z))^2].
              \end{aligned}
          \end{equation}
    \item $\ln f(z)$ middle
          \begin{equation}\label{eq}
              \begin{aligned}
                  \text{Re}\sigma(z+\Delta z)= & \text{Re}\sigma(z)+\frac{(\ln f(z-\Delta z)-\ln f(z+\Delta z))}{2}\text{Re}\sigma(z)+\Delta z\big[ 2\omega \text{Im}\sigma(z)\text{Re}\sigma(z)\big], \\
                  \text{Im}\sigma(z+\Delta z)= & \text{Im}\sigma(z)+\frac{(\ln f(z-\Delta z)-\ln f(z+\Delta z))}{2}\text{Im}\sigma(z)+\Delta z[\frac{\omega}{f^2(z)}-\frac{\mu^2z^2}{\omega f(z)}     \\&+\omega (\text{Im}\sigma(z))^2-\omega(\text{Re}\sigma(z))^2].
              \end{aligned}
          \end{equation}
    \item $f(z)$ forward
          \begin{equation}\label{eq}
              \begin{aligned}
                  \text{Re}\sigma(z+\Delta z)= & \text{Re}\sigma(z)+(1-\frac{f(z+\Delta z)}{f(z)})\text{Re}\sigma(z)+\Delta z\big[ 2\omega \text{Im}\sigma(z)\text{Re}\sigma(z)\big], \\
                  \text{Im}\sigma(z+\Delta z)= & \text{Im}\sigma(z)+(1-\frac{f(z+\Delta z)}{f(z)})\text{Im}\sigma(z)+\Delta z[\frac{\omega}{f^2(z)}-\frac{\mu^2z^2}{\omega f(z)}     \\&+\omega (\text{Im}\sigma(z))^2-\omega(\text{Re}\sigma(z))^2].
              \end{aligned}
          \end{equation}
    \item $f(z)$ middle
          \begin{equation}\label{eq}
              \begin{aligned}
                  \text{Re}\sigma(z+\Delta z)= & \text{Re}\sigma(z)+\frac{f(z-\Delta z)-f(z+\Delta z)}{2f(z)}\text{Re}\sigma(z)+\Delta z\big[ 2\omega \text{Im}\sigma(z)\text{Re}\sigma(z)\big], \\
                  \text{Im}\sigma(z+\Delta z)= & \text{Im}\sigma(z)+\frac{f(z-\Delta z)-f(z+\Delta z)}{2f(z)}\text{Im}\sigma(z)+\Delta z[\frac{\omega}{f^2(z)}-\frac{\mu^2z^2}{\omega f(z)}     \\&+\omega (\text{Im}\sigma(z))^2-\omega(\text{Re}\sigma(z))^2].
              \end{aligned}
          \end{equation}
\end{enumerate}

We remark that for both of the middle methods the first layer is not defined since the data of the current layer depends on the data in the previous and next layers. Thus, in practice we adopt the forward difference method to define the first layer. For more details, please see the appendix \ref{Appendix: The python code}.

\section{DNN training parameters and results}\label{Appendix: DNN training parameters and results}

In this appendix we present the details of the training process, including the setup for epochs, loss functions, learning rate, the optimization algorithm as well as the training criteria.

As a whole, the training process is divided into two steps. The first step contains 3001 epochs, while the second step contains 2001 epochs. The loss function for each step has been shown in the main body of the paper. 
The learning speed with $L_1$-loss is faster while $L_2$-loss can make the final metric more smooth.

For the optimization algorithm, we use the RMSprop optimizer in the first step and the Adam optimizer in the second step. The batch size is fixed as 200.

For the learning rate, we reduce it gradually along with the increase of the epoch by applying the module torch.optim.lr\_scheduler.MultiStepLR in Pytorch. At the first step, when the number of epochs is within (1,1000), the learning rate is set as $10^{-3}$. In the range of (1001,1500) and (1501,3001), the learning rate is set as $9\times 10^{-4}$ and $8.1\times10^{-5}$ respectively. Similarly, at the second step  when the number of epoch is within (1,500), (501-1000), (1001-1250), (1251,1500), (1501,1750) and (1751,2001), the learning rate is $10^{-3}$, $10^{-4}$, $10^{-5}$, $10^{-6}$, $10^{-7}$ and  $10^{-8}$ respectively. In principle, this kind of selection for learning rate is empirical and the epoch is large enough such that the loss will not reduce and fluctuate at some quantity.

The values of hyper-parameters $c_1$ and $c_2$ are specified quite casually because of our special design of the regularization term. In practice, we choose 50, 100 and 200 for $c_2$ and 1, 10 and 20 for $c_1$ correspondingly. During the training process, the harder is the training task, the larger the value of $c_1$ needs to be. Nevertheless, in general the setup of $c_1$ and $c_2$ does not affect the training results much. In particular, the involvement of the second training step  makes the specification of $c_1$ and $c_2$ less important.

The training criteria($MAE$ and $MSE$) are shown below,

\begin{equation}\label{eqscdc}
    \begin{aligned}
        MAE&=\sum_{i=1}^N\frac{|f^i_p-f^i_t|}{N}, \\
        MSE&=\sum_{i=1}^N\frac{(f^i_p-f^i_t)^2}{N},
    \end{aligned}   
\end{equation}
where $i$ represents the $i^{th}$ layer and $f^i_p$ is the quantity that the network trains, which is just the metric function $f(z)$ in this work. $f^i_p$  represents the metric of prediction, while $f^i_t$ refers to the true metric.

To prevent the influence of contingency factors and statistical fluctuations on the learning process, we train 5 times for each training process and set the average of these results as our final results. All the training results are listed as below (in the next page):
\begin{sidewaystable}
        \caption{All data of the black hole metrics training results}
        \label{data}
        \tabcolsep=0.3cm
        \renewcommand\arraystretch{0.8}
        \begin{tabular}{ccccccccccccc}
        \toprule [1pt]
            Number & Type    & $\mu$ & $z_h$ & difference type  & $\omega$     & $c_1$ & $c_2$ & $MAE_p$   & $MSE_p$   & $MAE_f$   & $MSE_f$   \\
            1      & real    & 1     & 0.9   & ln(f(z)) forward & (0.1,1{]}    & 1     & 50    & 2.856E-03 & 1.306E-05 & 1.519E-03 & 3.596E-06 \\
            2      & real    & 1     & 0.9   & ln(f(z)) forward & (0.99,1{]}   & 1     & 50    & 2.724E-03 & 1.213E-05 & 2.457E-03 & 1.014E-05 \\
            3      & real    & 1     & 0.9   & ln(f(z)) forward & (0.1,0.11{]} & 10    & 50    & 3.313E-03 & 1.570E-05 & 3.296E-03 & 1.627E-05 \\
            4      & real    & 1     & 0.99  & ln(f(z)) forward & (0.1,1{]}    & 1     & 50    & 3.205E-03 & 2.422E-05 & 2.278E-03 & 9.641E-06 \\
            \textcolor{blue}{5}       & \textcolor{blue}{real}  &\textcolor{blue}{1}   & \textcolor{blue}{0.99}            &\textcolor{blue}{f(z) forward}     &\textcolor{blue}{(0.1,1}{\textcolor{blue}{]}}  &\textcolor{blue}{1}   &\textcolor{blue}{50} &\textcolor{blue}{ 4.599E-02             }&\textcolor{blue}{3.235E-03}              & \textcolor{blue}{5.121E-02}       & \textcolor{blue}{4.516E-03}  \\
            6      & real    & 1     & 0.99  & ln(f(z)) forward & (0.99,1{]}   & 10    & 100   & 1.956E-02 & 6.080E-04 & 7.725E-03 & 1.128E-04 \\
            7      & real    & 1     & 0.99  & ln(f(z)) forward & (0.1,0.11{]} & 20    & 100   & 1.564E-02 & 3.622E-04 & 1.664E-02 & 3.632E-04 \\
            8      & reduced & 1     & 0.9   & f(z) forward     & (0.1,1{]}    & 1     & 50    & 3.485E-03 & 1.894E-05 & 2.564E-04 & 1.411E-07 \\
            9      & reduced & 1     & 0.9   & f(z) forward     & (0.99,1{]}   & 1     & 50    & 5.942E-03 & 5.149E-05 & 4.782E-03 & 3.399E-05 \\
            10      & reduced & 1     & 0.9   & f(z) forward     & (0.1,0.11{]} & 1     & 50    & 9.860E-03 & 1.377E-04 & 9.216E-03 & 1.273E-04 \\
            11     & reduced & 1     & 0.99  & f(z) forward     & (0.1,1{]}    & 1     & 50    & 2.208E-03 & 7.432E-06 & 2.042E-03 & 6.681E-06 \\
            \textcolor{blue}{12}       & \textcolor{blue}{reduced}  &\textcolor{blue}{1}   & \textcolor{blue}{0.99}            &\textcolor{blue}{f(z) middle}     &\textcolor{blue}{(0.1,1}{\textcolor{blue}{]}}  &\textcolor{blue}{1}   &\textcolor{blue}{50} &\textcolor{blue}{ 1.668E-02             }&\textcolor{blue}{9.275E-04}              & \textcolor{blue}{2.212E-02}       & \textcolor{blue}{1.175E-03}  \\ 
            13     & reduced & 1     & 0.99  & f(z) forward     & (0.99,1{]}   & 10    & 200   & 1.377E-02 & 3.220E-04 & 1.383E-02 & 2.970E-04 \\
            14     & reduced & 1     & 0.99  & f(z) forward     & (0.1,0.11{]} & 10    & 200   & 1.203E-02 & 2.009E-04 & 1.076E-02 & 1.526E-04 \\
            15     & real    & 2     & 0.9   & ln(f(z)) forward & (0.1,1{]}    & 1     & 50    & 4.261E-03 & 2.773E-05 & 1.661E-03 & 5.128E-06 \\
            16     & real    & 2     & 0.9   & ln(f(z)) forward & (0.99,1{]}   & 10    & 200   & 3.718E-02 & 2.372E-03 & 2.521E-02 & 1.070E-03 \\
            17     & real    & 2     & 0.9   & ln(f(z)) forward & (0.1,0.11{]} & 10    & 100   & 6.360E-02 & 5.514E-03 & 6.351E-02 & 5.479E-03 \\
            18     & real    & 2     & 0.99  & ln(f(z)) forward & (0.1,1{]}    & 1     & 200   & 1.036E-02 & 2.017E-04 & 4.278E-03 & 3.327E-05 \\
            19     & real    & 2     & 0.99  & ln(f(z)) forward & (0.99,1{]}   & 10    & 200   & 4.773E-02 & 4.332E-03 & 5.715E-02 & 6.621E-03 \\
            20     & real    & 2     & 0.99  & ln(f(z)) forward & (0.1,0.11{]} & 10    & 200   & 4.973E-02 & 3.772E-03 & 4.459E-02 & 3.045E-03 \\
            21     & reduced & 2     & 0.9   & f(z) forward     & (0.1,1{]}    & 1     & 50    & 3.292E-03 & 1.736E-05 & 2.139E-03 & 1.041E-05 \\
            22     & reduced & 2     & 0.9   & f(z) forward     & (0.99,1{]}   & 10    & 100   & 4.151E-02 & 3.483E-03 & 3.457E-02 & 2.473E-03 \\
            23     & reduced & 2     & 0.9   & f(z) forward     & (0.1,0.11{]} & 10    & 100   & 5.500E-02 & 5.048E-03 & 5.534E-02 & 5.152E-03 \\
            24     & reduced & 2     & 0.99  & f(z) forward     & (0.1,1{]}    & 1     & 50    & 2.860E-03 & 1.336E-05 & 2.779E-03 & 1.323E-05 \\
            25     & reduced & 2     & 0.99  & f(z) forward     & (0.99,1{]}   & 10    & 100   & 6.015E-02 & 6.861E-03 & 4.822E-02 & 4.326E-03 \\
            26     & reduced & 2     & 0.99  & f(z) forward     & (0.1,0.11{]} & 10    & 200   & 8.112E-02 & 1.012E-02 & 8.092E-02 & 1.013E-02\\
            \toprule [1pt]
        \end{tabular}
        
\end{sidewaystable}

 \newpage

\section{The python code}\label{Appendix: The python code}
We employ PyTorch (CPU version 1.6.0) and Python (version 3.7.9) to implement the deep learning process in this work. It is suggested to run this in Anaconda3. The modules such as PyTorch, torchvision, and moviepy should be installed in advance. The main code and a set of training data can be found in \cite{dataset}.

\end{document}